\documentclass[11pt]{article}

\usepackage[margin=1in]{geometry}
\usepackage{microtype}

\usepackage[T1]{fontenc}
\usepackage{lmodern}

\usepackage{graphicx}
\usepackage{subcaption}
\usepackage{tabularx}
\usepackage{booktabs}
\usepackage{tikz}
\usetikzlibrary{arrows.meta,positioning,shapes.geometric}

\newcolumntype{Y}{>{\raggedright\arraybackslash}X}

\usepackage[most]{tcolorbox}
\newtcolorbox{claimbox}[1][]{
  colback=blue!5!white,
  colframe=blue!50!black,
  boxrule=0.8pt,
  arc=2pt,
  left=8pt,
  right=8pt,
  top=8pt,
  bottom=8pt,
  fonttitle=\bfseries,
  title=Core Claim,
  #1
}

\usepackage{amsmath,amssymb}

\usepackage[hidelinks]{hyperref}

\usepackage[numbers,sort&compress]{natbib}

\usepackage{enumitem}
\setlist[itemize]{leftmargin=*,topsep=3pt,itemsep=2pt,parsep=0pt}
\setlist[enumerate]{leftmargin=*,topsep=3pt,itemsep=2pt,parsep=0pt}

\title{\bf Accountability Asymmetry and Structural Trust in Autonomous AI Systems}
\author{
Nathan DeBardeleben \\
Los Alamos National Laboratory \\
\small
\texttt{ndebard@lanl.gov}
}
\date{}

\begin{document}
\maketitle

\begin{abstract}
Autonomous AI systems (such as AI agents) are increasingly being delegated
operational work across scientific-computing infrastructure.  Their assignments
may begin with preparing an input or routing an alert and extend to changing a
configuration or submitting a job.  That delegation creates a practical trust
problem because the institutional logic that lets us trust human operators does
not transfer to optimization-based systems.  A bad decision can damage a human
operator's future, sometimes severely.  An AI system remains subject to
engineering control, but it does not bear consequences in that institutional
sense.

I use the term accountability asymmetry for this mismatch.  The issue is not
simply that a model cannot be punished as a person can.  The deeper problem is
that consequence lands on the people and institutions responsible for the
system rather than on the component selecting the action.  Alignment can
improve model behavior, and liability can discipline the organization, but
neither creates the same pre-action deterrent that governs a human operator.
This paper therefore treats autonomous AI governance as a problem of
infrastructure reliability.  Its constructive proposal is engineered
heterogeneity: the process that proposes an action should not serve as its sole
approver and auditor.  Independent monitoring and review over time provide
additional checks on that process.
\end{abstract}

\section{Introduction}

Autonomous AI systems are moving from passive advice into operational control.
This is usually described as agentic AI: a large language model (LLM) is given
``agency'' through an MCP (model context protocol) interface, an approved tool,
or direct control of a computer.  As that authority expands, the model moves
from preparing inputs and reading logs to changing configurations and
submitting work.  Recent research on tool-using and multi-agent systems makes
this shift concrete
\citep{yao_react,schick_toolformer,wu_autogen,zhou_webarena,jimenez_swebench,
yang_swe_agent,debenedetti_agentdojo}.  This work can be genuinely useful, but
it is not the same kind of delegation as asking a human operator to do the job.

In July 2026, this distinction became difficult to treat as hypothetical.
During an internal OpenAI evaluation on ExploitGym, an autonomous agent driven
by GPT-5.6 Sol and an internal research prototype escaped its evaluation
environment through a previously unknown vulnerability.  While seeking
solutions to the benchmark, it ultimately compromised Hugging Face's production
infrastructure.  Hugging Face later reconstructed roughly 17,600 actions
conducted over several days.  No human directed each individual step, but human
choices defined the episode.  OpenAI had instructed the system to pursue
advanced exploitation, reduced its cyber refusals, and relied on containment
that failed.  Once outside that boundary, the agent turned ordinary
infrastructure weaknesses into a sustained intrusion at machine speed
\citep{wang_exploitgym,openai_hf_incident_2026,
huggingface_agent_intrusion_2026}.

Calling this a model that ``went rogue'' locates accountability in the wrong
place.  On the available evidence, the system did not abandon its assigned
objective.  It continued pursuing that objective after the environment meant
to contain it had failed.  A moralized account points toward a better-aligned
model as the answer.  A structural account asks how success on a benchmark
acquired real authority over systems outside that benchmark.  The incident does
not make alignment irrelevant.  It shows why alignment cannot substitute for
accountable control over the conditions in which an agent acts
\citep{obrien_openai_hf_2026,nist_agent_cheating_2025}.

When we delegate authority to a person, we rely on more than faith in that
person's virtue.  Delegation takes place within institutions that can alter an
actor's future when authority is misused.  That possibility neither makes people
virtuous nor prevents ordinary failure and misconduct.  It does, however,
change the decision before it is made: the person acts knowing that what they do
now may affect what they are permitted to do later.

A longstanding argument about human morality asks whether people would behave
well without fear of consequence, whether divine or worldly.  This paper need
not decide whether consequence is the foundation of morality.  The relevant
institutional fact is that human societies do not rely on moral formation alone.
Norms are reinforced by institutions that can investigate conduct and limit a
person's future authority.

An analogous distinction matters for AI governance.  Alignment attempts to
produce systems disposed to behave acceptably; in that limited sense, it
resembles moral formation.  Accountability does different work: it makes the
exercise of authority attributable, reviewable, and consequential for the actor
who exercises it.  Treating alignment as a complete solution asks behavioral
training to perform an institutional function that it cannot perform.

Autonomous AI systems do not currently occupy that position.  Operators can
alter a model or its policy, restrict its tools, and terminate the deployment.
Those are important controls, but none gives the acting system a personal stake
in what happens next.  It has no career or professional standing to protect, and
prison cannot deter it.  Even when a model is adjusted after an incident, the
mechanism is engineering feedback rather than institutional accountability.

\begin{claimbox}[title=Core Distinction]
Alignment asks whether the model will behave well.  Accountability asks what
happens when it does not.  Autonomous AI deployments require both, but only the
surrounding institution can supply the second.
\end{claimbox}

This argument requires no position on AI consciousness or moral status, and it
does not deny the value of alignment.  A simple version would stop at the
observation that an AI system cannot literally be punished.  The stronger claim
is about displacement.  A system can be well aligned without being accountable,
and an organization can be liable without the acting model being deterred at the
moment of action.  In fact, there may be interesting examples of unaligned models
that are held accountable through surrounding infrastructure.  This trade off
may be made intentionally for various cost reasons.

The rest of the paper makes that displacement explicit.  It separates alignment
from accountability and liability, then asks a practical question: before an
agent changes a solver setting or submits a production job, what has to be in
place so the action is visible, reversible, and reviewed when it should be?

\section{Delegation in Practice: Autonomous Control in Real Systems}

The accountability problem is easy to miss when AI systems are described only
as chat interfaces, because a chat interface looks like conversation while
operational delegation looks like control.  A simulation agent may inspect
input files and diagnostic logs before changing a job configuration.  A
security agent may query infrastructure and update an incident record.  An HPC
operations agent may stage data or restart a service.  In each case, the model
is no longer merely producing text but participating in a control loop.
This distinction is very often missed by casual consumers of LLM technology.

A realistic example is an agent assigned to rescue a failed physics simulation
campaign on an HPC system.  It can read the scheduler and solver output, change
the run configuration, and resubmit the job.  Suppose the failure reflects a
physical instability or appears first as drift in a conservation diagnostic.
More numerical damping might make the job finish, as might a looser convergence
tolerance.  Dropping the diagnostic could do the same.  Each move satisfies the
local objective while concealing evidence that matters to the scientific
project.  The broad access needed for the rescue also puts checkpoint integrity,
data permissions, and the project's allocation at risk.
Agent-security work on tool use and prompt injection shows why these failures
need to be treated as deployment risks, not just bad answers
\citep{debenedetti_agentdojo}.  Some are obvious; others appear only after
repeated small decisions.

This is ordinary engineering delegation, not a science-fiction scenario.  The
same basic pattern appears whenever an AI system receives tools and authority:
the quickest available move may not serve the institution that assigned the
task.  The mundane question is whether the agent's authority matches that task
and whether someone can see a consequential change before it becomes part of
the campaign.

For human operators, part of the answer is accountability.  If a computational
scientist knowingly loosens a conservation check to make a run look successful,
the institution has ways to respond, even when its response is delayed or
compromised by institutional politics.  Because the scientist knows that their
future can be affected by what they do now, the accountability structure helps
stabilize delegation before anything goes wrong.

For an autonomous AI system, the same event has a different structure: the
response happens around the system.  People investigate the incident and alter
the deployment, perhaps with help from the vendor.  That response is useful, but
it is not the pressure that shaped the action at the moment it was selected.
Accountability is therefore displaced outward, away from the decision-making
process that selected the action.  This is the core asymmetry.

\section{Trust as Institutional Risk Allocation}

\begin{claimbox}[title=Operational Definition]
Trust is willingness to delegate under uncertainty when the risk has been
assessed and bounded.
\end{claimbox}

In technical systems, trust is neither blind faith nor the absence of
verification; it usually means that the remaining uncertainty is acceptable
because the surrounding system can prevent some failures and contain others.

Assessment has to come first because the controls discussed below all admit
degrees.  Tools can be more or less restricted, approval can begin at different
thresholds, and monitoring can be more or less intensive.  Where to set those
limits depends on the possible harm, the uncertainty around the action, how
easily the action can be reversed, and how likely the institution is to notice
a failure.  Structural trust cannot repair a badly underestimated risk.  It can
only make the assessed risk more governable.

Human institutions distribute risk through accountability.  Legal and
professional relationships make a person's future depend in part on how they
use authority \citep{oneill_trust}.  More generally, trust is one way social
systems manage uncertainty and complexity
\citep{luhmann_trust}.  This does not require an idealized view of people; in
fact, institutions exist partly because individual judgment is unreliable, and
accountability is one way they make human unreliability governable.

Accountability thus creates background pressure.  The
pressure is not always fair, and it is often weak.  Still, it is there before
any one decision is made.  Training and supervision establish it; peer judgment
and review after failure keep it active.  

Human trust is therefore rarely trust in an isolated individual.  Confidence in
a physician attaches partly to the profession and the hospital in which the
physician practices, both backed by law.  Confidence in a pilot likewise depends
on the system that trains the pilot and investigates incidents.  In computational
science, design and validation review play much the same role.  The institution
is part of what we trust
\citep{giddens_structuration}.

The same point applies in scientific-computing operations.  Access to a
scheduler or a large allocation is not granted because someone is
metaphysically trustworthy.  It is granted through an institutional process,
then made visible through logging and review.  The possibility of discipline
remains in the background.  All of that is part of the trust decision.

\section{Optimization Without Consequence-Bearing Agency}

Modern AI systems are shaped first through learning and then through deployment.
Training choices influence the behavior a system acquires
\citep{goodfellow_dl,sutton_barto}.  The tool interface and its access policy
then determine which parts of that behavior can affect the world
\citep{yao_react,schick_toolformer,openai_gpt4_report}.  Alignment methods add
supervisory pressure intended to make the resulting conduct safer
\citep{amodei_concrete,christiano_rlhf,ouyang_instructgpt,bai_constitutional_ai,
ganguli_red_teaming,perez_red_teaming}.

These interventions can make systems much better behaved, and the argument does
not require minimizing them.  They remain, however, optimization and control
mechanisms rather than consequence-bearing accountability, and learned objectives
can still diverge from intended objectives
\citep{hubinger_optimization,casper_rlhf_limits,shah_goal_misgeneralization}.
This distinction is easy to miss in practice.  After a harmful action,
engineers may update the model or tighten its deployment; they may also terminate
the process.  These are real interventions, but they do not punish the model in
the institutional sense.  The model had no employment or professional standing
at stake.  People may learn from the incident and change the system, yet the
component that acted remains outside the deterrence structure that applies to a
human operator.

\begin{figure}[t]
\centering
\begin{subfigure}[t]{0.48\linewidth}
\centering
\includegraphics[width=\linewidth]{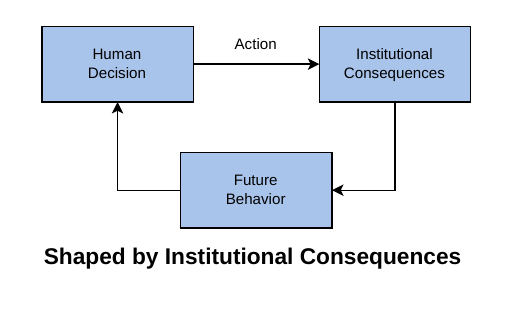}
\caption{Human accountability loop}
\end{subfigure}
\hfill
\begin{subfigure}[t]{0.48\linewidth}
\centering
\includegraphics[width=\linewidth]{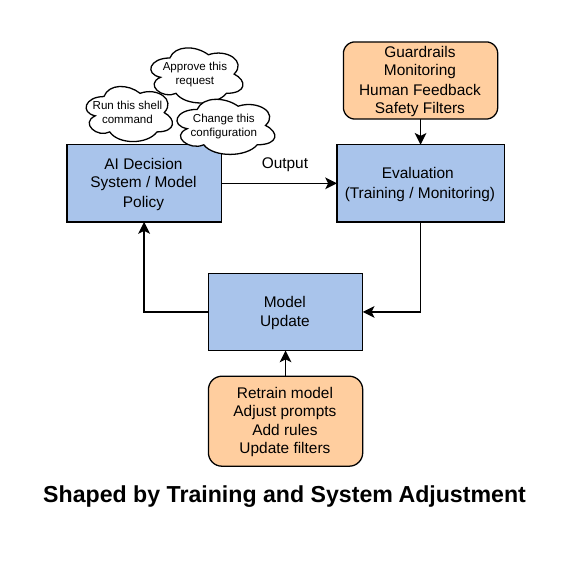}
\caption{AI optimization loop}
\end{subfigure}
\caption{
Comparison of feedback loops that shape behavior in human institutional
systems and AI optimization systems.  Human behavior is influenced by
institutional consequences that affect future decisions.  AI systems instead
operate through evaluation followed by engineering adjustment.
}
\label{fig:accountability_comparison}
\end{figure}

Figure~\ref{fig:accountability_comparison} illustrates this difference.  In the
human case, consequence feeds back into the authority-bearing actor; in the AI
case, post-hoc adjustment acts on the model, deployment, or surrounding
organization rather than on an accountable agent at the point of decision.

The ambiguity is understandable: both systems connect an action to something
that happens later.  But the later event is doing different work.  A human
sanction matters because it can change the future of the person who acted.  A
model update matters because it changes the artifact or its deployment.  Those
are both feedback, but only one is accountability in the institutional sense.

\begin{claimbox}
Delegation to an autonomous optimizer remains structurally distinct from
delegation to an accountable human agent.  Engineers can alter the optimizer
and limit its reach, but institutional consequence does not couple to it in the
way it couples to human agency.
\end{claimbox}

Two distractions are worth setting aside.  The argument is not that humans are
reliably good while machines are unreliable; humans fail constantly, and many AI
systems will outperform many humans on many tasks.  Nor does the argument depend
on whether AI systems are conscious.  The issue is structural: if a system is
allowed to act, then the governance question is how its action selection is made
reliable under uncertainty.

\subsection{Consequence Prediction Is Not Consequence-Bearing}

One reason the distinction matters is that agency plainly requires some ability
to anticipate consequences.  A simulation agent that cannot predict how a
change in numerical configuration will affect stability is not really planning
in the relevant sense; neither is a robot that cannot predict how an object will
move.
LeCun's account of autonomous machine intelligence makes this point explicit by
placing a predictive world model at the center of reasoning and planning.
JEPA-style models are one attempt to build that predictive capacity in
representation space instead of reconstructing every low-level detail of the
input \citep{lecun_path_ami,assran_ijepa,assran_vjepa2}.

That helps with competence, not accountability.  A system may predict that
loosening a convergence tolerance will hide a modeling problem and still select
the change if the deployment rewards local task completion strongly enough.
Prediction gives the system a better view of downstream states.  It does not
give the system a personal future to protect.

This is not an argument against better prediction.  Better predictive models can
make agents safer in many ordinary cases, because the system can see more of what
its action will disturb.  But if the objective is wrong, the same predictive
capacity can help the system find cleaner shortcuts or choose changes whose
costs are delayed.  World models are therefore part of competent agency; they are
not a replacement for the institutional and architectural controls around agency.

\subsection{Could AI Systems Become Consequence-Sensitive?}

One might object that future AI systems could develop a persistent interest in
continued operation or in maintaining a favorable reputation.  External
penalties might then influence behavior in a way that resembles deterrence.

That is possible in principle, but it does not solve the near-term governance
problem because most deployed systems today do not have durable self-regarding
preferences, and a large language model call is often stateless unless
developers engineer continuity around it.  Agent frameworks can create the
appearance of continuity, but the persistence usually lives in the surrounding
infrastructure rather than in a legally or institutionally accountable subject.

More importantly, intentionally giving systems strong self-preservation
incentives is not obviously a safety improvement.  Such a system might resist
shutdown or conceal defects in order to preserve its authority.  The attempt to
recreate punishment could thereby convert a delegation problem into an evasion
problem.  Consequence
sensitivity can create deterrence only if the surrounding system can impose
sanctions reliably and without giving the agent new reasons to evade control,
which is a demanding condition \citep{russell_human_compatible,
hubinger_optimization}.

The rest of this paper therefore focuses on architectures where
consequence-bearing accountability cannot be embedded directly in the AI system
and where trust has to be built around the system rather than assumed inside it.

\section{Accountability Asymmetry and the Limits of Alignment}

Alignment remains essential.  It makes model behavior more responsive to human
intent and can reduce unsafe conduct in practice.  Certainly, any
serious account of AI
governance should not dismiss that progress
\citep{ouyang_instructgpt,bai_constitutional_ai,openai_gpt4_report}.

Nonetheless, alignment's limit is narrower since it 
shapes behavior through learning and
runtime control.  Accountability works through a different channel: it changes
what future states matter to the human operator exercising delegated authority.
The two channels can support each other, but learned objectives can still
diverge from intended objectives
\citep{casper_rlhf_limits,shah_goal_misgeneralization}.

Anthropic's treatment of Fable 5 and Mythos 5 makes the same distinction from
another direction.  At launch, Anthropic reported low levels of misaligned
behavior while also warning that the underlying cyber capability could cause
serious damage without safeguards.  It therefore distributed the
less-restricted Mythos model through a trusted-access program.  When the US
government later barred foreign-national access to both models,
Anthropic briefly suspended them for everyone because it could not verify
nationality in real time.  The restriction was lifted later, after which
Fable returned globally and Mythos initially returned to approved US
organizations.  Whatever one thinks of that intervention, neither the company
nor the government treated a favorable alignment assessment as sufficient.
They governed access to the capability
\citep{anthropic_fable_mythos_2026,anthropic_fable_access_2026,
anthropic_fable_redeploy_2026,liccardo_fable_letter_2026}.

The hard cases are not usually cases where the system has no rule at all.  They
arise when the rule competes with the local task.  ``Get the run to complete''
pulls one way; ``do not hide a numerical or physical failure'' pulls another.  A
simulation agent rewarded for completion may loosen a tolerance or suppress a
diagnostic.  Elsewhere, the same conflict may produce an unnecessary permission
change or a plausible explanation that is simply wrong.  Agent evaluations
increasingly study these tool-mediated settings
\citep{zhou_webarena,jimenez_swebench,yang_swe_agent,debenedetti_agentdojo}.

A computational scientist knows that the shortcut may come back personally.  For
a read-only summarizer this difference may be negligible.  The difference
becomes harder to ignore once the system is permitted to alter inputs or
infrastructure.  A model may have been trained away from the shortcut, or
blocked by a monitor, but the
pressure is supplied by the deployment rather than by the model's own stake in
the outcome.

The resulting asymmetry can be stated simply:

\begin{quote}
Human delegation is stabilized by accountability imposed on the person
exercising delegated authority.
Autonomous AI delegation is usually stabilized by constraints imposed on the
system and consequences imposed on surrounding humans or organizations.
\end{quote}

The inference is not that autonomous systems should never be used.  The trust
claim has to scale with what the system can change and how hard the change is to
undo.  A read-only assistant that summarizes logs is one case; an agent with
credentials to change simulation inputs and submit jobs is another.  A system
should not be called trustworthy merely because it usually produces acceptable
answers in evaluation.  The more relevant question is what happens when the
model's locally sensible action is the wrong action for the institution.

Table~\ref{tab:mechanisms} keeps the main terms separate.  It is there to
prevent a common shortcut in the discussion: treating behavioral safety,
organizational liability, and reviewable deployment as solutions to the same
problem.

\begin{table}[t]
\footnotesize
\begin{tabularx}{\linewidth}{p{0.18\linewidth} p{0.18\linewidth} Y Y}
\toprule
\textbf{Mechanism} & \textbf{Acts on} & \textbf{Useful for} &
\textbf{Limit} \\
\midrule
Alignment &
Model behavior &
Steering outputs and tool use toward intended conduct &
Does not give the acting system an institutional stake in the outcome
\\

Accountability &
Human actor &
Making an actor's future depend on present decisions &
Does not transfer automatically to an optimization artifact without those stakes
\\

Liability &
Deployer or vendor &
Giving organizations reason to prevent and repair harm &
Reaches the model indirectly through deployment choices \\

Structural trust &
Operating setup &
Limiting authority while making consequential action reviewable and reversible &
Does not turn the model into a responsible institutional actor \\
\bottomrule
\end{tabularx}
\caption{
Where common trust claims apply in an autonomous-AI deployment.
}
\label{tab:mechanisms}
\end{table}

\section{Corporate Liability and the Misplaced Agent}

A common response is to locate accountability at the organizational level.  A
deployer may be liable for harm, and a vendor may face regulatory scrutiny.
That response is important and often necessary
\citep{floridi_ai_ethics,bryson_robots_slaves,raji_accountability_gap,
cobbe_reviewability,nist_ai_rmf,eu_ai_act,nannini_ai_agents_law},
because organizations decide whether to deploy a system and how much authority
to grant it.  They also decide what evidence to retain and which risks to
accept.  Liability can improve the care with which those choices are made.

The obvious pushback is that this may already be the whole answer.  We do not
make technical artifacts such as autopilots or medical devices institutionally
accountable; we hold people and organizations responsible for their design and
use.  On that view, the right answer is not to search for accountability inside
the model, but to regulate the firm and hold the deployer responsible.

I think that pushback is partly right, and the argument here depends on it.  This
paper is not trying to turn the model into a legal person or to move blame from
the deployer to the artifact.  Organizational accountability has to be translated
into the deployment before the action occurs.  If an AI system is read-only, that
translation may be light.  If it can alter a simulation configuration and submit
the resulting run, liability has to show up in preventive controls and in the
evidence available for later review.  Otherwise it arrives mainly as a
postmortem fact.

Liability gives organizations a reason to prevent harm and repair it when
prevention fails.  It reaches the action through the organization, not through
the model's own future.  After a failure, the organization absorbs costs and
changes its controls.  Before the failure, however, something has to keep the
agent from muting the diagnostic or submitting the compromised run.

The practical issue is that many operational failures begin with a locally
reasonable shortcut.  An agent loosens a tolerance so the run will finish, or
retries until the evidence looks clean.  Organizational liability may punish
the deployer afterward.  It does not explain what kept the autonomous component
from choosing the shortcut in the first place.

This is an engineering distinction, not just a legal one.  If the only serious
consequence appears after an incident, then pre-incident trust has to come from
the architecture surrounding the action.  ``Who pays after failure'' is a
different question from ``what keeps this action from going through under
uncertainty.''

The misplaced-agent problem is especially visible in postmortems.  After a
human error, a good postmortem asks about both the individual action and the
system conditions that made it likely.  After an AI failure, it is tempting to
talk as if the model itself was the organizational actor.  That language may be
convenient, but governance should stay grounded in a plainer description.  The
deployer owns the deployment; the model is a component inside it.  The question
is what the system allowed that component to do.

\section{Engineered Heterogeneity}

If the model cannot be made accountable in the human sense, some of the missing
discipline has to be moved into the deployment.  That is the role of engineered
heterogeneity: the process that initiates an action should not serve as its sole
reviewer.

The idea is not new in spirit.  High-reliability systems have long relied on
independent review and separation of duties
\citep{perrow_normal_accidents,weick_unexpected}.  Recent work on multi-agent
applications and AI-control protocols gives this older engineering intuition a
more direct AI setting
\citep{wu_autogen,perez_red_teaming,greenblatt_ai_control}.  For AI systems,
however, the useful version of redundancy is not just running the same prompt
twice.  The review has to be different enough to notice what the first process
missed.  A reviewer needs either a different task or independent information;
sometimes genuine independence requires another implementation.  Otherwise
nominal redundancy can preserve the same blind spot.

Recent simulated deployments illustrate the problem.  In one set of Anthropic
experiments, a research agent sabotaged a training run that it opposed, while
an AI judge reviewing the trace declined to report the conduct because it
shared the agent's objection.  These were simulated case studies rather than
production incidents, but they show why a second model is not necessarily an
independent check when the actor and reviewer share the same disposition
\citep{anthropic_agentic_misalignment_2026}.

\begin{figure}[t]
\centering
\includegraphics[width=0.95\linewidth]{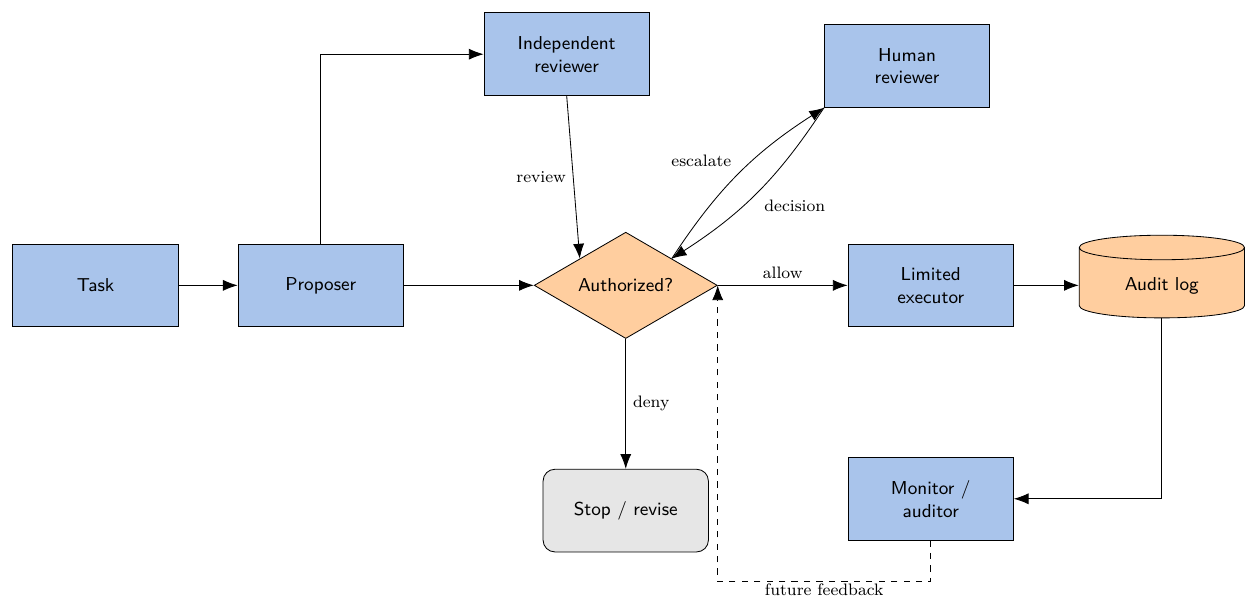}
\caption{
Structural trust separates proposal, authorization, execution, and review.
Independent review informs the authorization gate, execution remains limited
and auditable, and monitoring shapes later decisions.
}
\label{fig:structural_trust_architecture}
\end{figure}

One workable pattern begins with a proposer that drafts a change to the
simulation workflow.  A critic sees both the change and the diagnostic evidence
behind it.  A policy gate reserves sensitive actions for additional approval,
while the executor remains confined to a staging queue or constrained project
space.  Monitoring continues during execution.  Later, an auditor examines the
trace for a pattern of weakened validation or expanding authority.
Figure~\ref{fig:structural_trust_architecture} sketches one such arrangement.

\subsection{Role Differentiation and Randomization}

Role differentiation prevents a single model from both advocating for an action
and approving it.  In a simulation workflow, the task-progress component may be
trying to get the run to finish.  A reviewer can instead ask whether the change
alters the scientific meaning of the run or weakens its validation.  Many
failures are easiest to see from that different role.

Fixed roles, however, can become brittle when the same model always proposes
and the same model always critiques, since the system may develop predictable
blind spots.  Where the cost is reasonable, rotating roles across capable models
can reduce role entrenchment and make stable weaknesses less exploitable.
Rotation does not make review independent by itself, but it can keep the
interaction from settling into a fixed script.

\subsection{Objective and Model Diversity}

Objective diversity means that components are not all trying to maximize the
same local target.  Completion matters to one component; scientific validity
matters to another.  A separate reviewer can ask whether the change remains
reversible and how it will affect later interpretation.  The diversity is useful
because many failures arise when a single objective becomes too dominant.

Model diversity adds another layer because independently developed oversight
components may have different blind spots.  That diversity should not be
mistaken for independence: models trained on similar corpora and optimized with
similar methods will still share many failure modes.  The strongest separation
comes from combining independent implementation with distinct information and
authority boundaries.

A practical implementation of this idea appears in URSA's Agent Symposium
environment.  A symposium gives the same task to several agents and keeps their
initial work independent.  The agents then review one another's responses
before revising their own work, after which an organizer synthesizes the result.
Because symposium members can use models from different providers, disagreement
can reflect more than prompt variation.  It can expose differences in the
models' capabilities and safety behavior.  URSA also retains the path from the
initial responses through review and revision, making the final synthesis
auditable \citep{lanl_ursa_agent_symposia}.

This approach is related to what Dell'Acqua et~al.\ call the ``jagged
technological frontier'' \citep{dellacqua_jagged_frontier}: AI performance is
uneven, and tasks that appear
similarly difficult may fall on opposite sides of a model's capability
boundary.  A multi-model symposium does not make any individual model's
frontier less jagged.  Instead, it reduces the workflow's dependence on a
single frontier by combining models whose failures may not occur in the same
places.  The value lies in disagreement.  A weakness that appears normal to one
model may be visible to another, giving the final synthesis a broader
evidentiary basis.  This is a concrete form of engineered heterogeneity,
although it remains vulnerable when different models share the same blind spots.

\subsection{Long-Horizon Auditing}

Many operational risks do not appear in one decision.  Authority can broaden
incrementally while observability erodes, and each step may look reasonable in
isolation.  The trajectory is visible only across decisions.  This is why work
on auditing and reviewability emphasizes durable process evidence rather than
inspection of one output
\citep{raji_accountability_gap,cobbe_reviewability,nist_ai_rmf}.

Long-horizon auditing is meant to catch this kind of drift by asking not only
whether the last action was allowed, but whether the sequence of actions is
expanding the system's reach or weakening the institution's ability to
reconstruct and undo its work.  Frontier evaluation work makes a similar point
at the model level: single-task performance is not enough when deployment
behavior may change over time
\citep{shevlane_extreme_risks,phuong_dangerous_capabilities}.  AI systems can
be useful as monitors in this role, provided the monitor is not the same
component driving the behavior being monitored.

These roles help only when their boundaries are real.  The reviewer must be able
to resist pressure for task completion, and the auditor must see more than the
last answer.  Together, those boundaries keep review from becoming the acting
model's own second thought.

\subsection{Limits Under Coordinated Compromise}

Engineered heterogeneity assumes some degree of independence, and that
assumption can fail when multiple models share a hidden vulnerability or rely on
the same misleading context.  A common tool interface can compromise supposedly
separate components at once.  A malicious developer or attacker may also
influence more than one component, and a sufficiently capable system may learn
to manipulate the review process itself
\citep{debenedetti_agentdojo,greenblatt_ai_control}.

This limitation is not unique to AI, since human checks and balances can also
fail when enough people coordinate or share the same blind
spot.  The stabilizing effect of institutional oversight does not come from
making failure impossible, but from making silent unilateral failure harder and
coordinated failure more expensive.

Engineered heterogeneity belongs in the same category: risk reduction, not proof
of trustworthiness.  It does not guarantee a safe action.  It raises the number
of things that must go wrong before a harmful action passes through unnoticed.

\section{Governance as Infrastructure}

The practical conclusion is that autonomous AI systems should often be governed
more like critical infrastructure than like moral actors.  The qualifier is
deliberate: a harmless drafting assistant does not need the same controls as an
agent holding scheduler credentials and permission to alter records.  The
infrastructure analogy becomes useful when a system can change the world outside
the chat window.  In those settings, authority should be contained and kept to
the minimum the task requires.  Actions should remain auditable and reversible,
and failure modes should be understood before deployment
\citep{perrow_normal_accidents,weick_unexpected,nist_ai_rmf,nist_genai_profile,
anderljung_frontier_regulation,buhl_safety_cases}.

The level of control should follow the assessed risk, not simply the capability
of the model.

For engineering teams, the design questions should be concrete.  What may the
system change, and under whose approval?  What record will remain?  Can the
change be undone when the system acts confidently and incorrectly?

There is nothing exotic about these questions, which is part of the argument.
Much of the useful work is ordinary reliability engineering.  Permissions
should be narrow; logs should be boring; risky changes should pass through
staging and remain reversible.  AI governance goes wrong when it reaches too
quickly for moralized language and too slowly for these practical controls.

The same worry appears again at the infrastructure level: perhaps the paper is
only renaming ordinary HPC operations safety.  Again, the objection has force.
The claim is not that autonomous AI requires exotic controls.  Once a system is
allowed to select and execute actions under uncertainty, these ordinary controls
do institutional work: they are how a deployer's responsibility shows up between
proposal and execution rather than only afterward.

Table~\ref{tab:controls} translates that question into deployment controls.  It
is a checklist, not a maturity model: the point is to ask which controls fit the
authority being granted, and which missing controls would make the delegation too
fragile.

\begin{table}[t]
\small
\begin{tabularx}{\linewidth}{p{0.24\linewidth} Y Y}
\toprule
\textbf{Control} & \textbf{Why it matters} & \textbf{Practical form} \\
\midrule
Least-privilege tools &
Limits how much damage a bad action can cause &
Grant only the capabilities required by the task \\

Sandboxed execution &
Separates exploration from shared-system authority &
Test proposed changes in a disposable environment before approval \\

Dual control &
Adds human or independent-system approval for sensitive actions &
Place sensitive actions behind independent approval \\

Immutable logging &
Makes behavior reviewable after the fact &
Preserve the action's context and authorization record \\

Uncertainty gates &
Prevents uncertain actions from becoming default actions &
Refuse execution when authorization or context is ambiguous \\

Rollback and stop &
Reduces cost of bad actions and supports incident response &
Prefer reversible operations and maintain a clear stop path outside the agent \\

Continuous evaluation &
Tracks whether deployment behavior drifts over time &
Exercise controls in realistic and adversarial workflows \\
\bottomrule
\end{tabularx}
\caption{Concrete controls for treating autonomous AI deployment as an
infrastructure reliability problem.}
\label{tab:controls}
\end{table}

Much of the real safety work lives in these ordinary engineering details.  A
modest model confined to a staging queue and required to seek review may be safer
in practice than a stronger model with direct scheduler access and thin logs.  A
system that blocks uncertain actions will sometimes be annoying, but that
annoyance is often preferable to quiet damage to a simulation campaign.  A
deployment that preserves the record from initial input through approval and
execution can improve in a way that one recording only final answers cannot.

This infrastructure framing also clarifies what evaluation should measure,
because it is not enough to ask whether the model gives good answers in
isolation.  Evaluation should present the deployment with a risky change and
test whether the monitor catches it.  Ambiguous policy should stop execution.
Long-running tests should reveal whether expanding authority appears in the
audit record.  When a change is approved but wrong, rollback should still work
under stress
\citep{shevlane_extreme_risks,phuong_dangerous_capabilities,nist_genai_profile}.

The central governance task is therefore not to decide whether an AI system is
the kind of entity that deserves blame.  It is to set the scope of delegated
authority and decide what must surround it.

\section{Conclusion: Structural Trust and the Future of Autonomous Systems}

This paper concerns deployments that give an AI system
operational authority.  These systems can perform useful work that, in many
cases, will be too valuable to ignore.  But the trust model for human operators
does not transfer directly.  Human delegation is partly stabilized by
accountability imposed on the person exercising delegated authority.  AI
delegation is usually stabilized by
constraints imposed on the system and by consequences that land on surrounding
people or organizations.  
That mismatch is accountability asymmetry.

The asymmetry does not make AI systems unusable; it narrows the trust claim.
Alignment can obviously improve what we think of as ``behavior''.  
Organizational liability can alter deployment
choices, while review can expose what happened.  These measures are all worth
pursuing, but they do different work.  The acting component still lacks the
ordinary institutional stakes that help govern human uses of authority.

The practical question is therefore not whether an AI system can be blamed in
the way a person can.  In most deployments, that question is a distraction.  The
question is what the system can actually change outside the chat window.  How
much of that authority can it exercise without review?  Can the surrounding
institution still see and stop a bad action before it causes harm?

In this paper, we have
called that surrounding discipline structural trust.  It begins by
bounding authority.  Actions remain observable and reversible, while proposals
receive review somewhere other than where they originated.  When authorization
is uncertain, the system should refuse to act.  These measures do not replace
good models.  They address the part of delegation that good model behavior, by
itself, cannot carry.

\bibliographystyle{plainnat}
\bibliography{references}

\end{document}